\documentstyle[prd,preprint,aps]{revtex}
\tightenlines

\begin{document}
\parskip 0.3cm
\begin{titlepage}

\begin{centering}
{\large \bf Signals for Majorana neutrinos in a $\gamma \gamma$ collider}\\
\vspace{.9cm}
{\bf J.Peressutti and O.A.Sampayo}\\
\vspace{.05in}
{\it  Departamento de F\'{\i}sica,
Universidad Nacional de Mar del Plata \\
Funes 3350, (7600) Mar del Plata, Argentina} \\ \vspace{.4cm}
 \vspace{.05in}
{ \bf Abstract}
\\
\bigskip
\end{centering}
{\small
 We study the possibilities to detect Majorana neutrinos in
 $\gamma \gamma$ colliders for different center of mass energies.
 We study the $W^\pm W^\pm l_j^{\mp} l_k^{\mp}$ ($l_j\equiv e ,\mu ,\tau$)
 final states which are, due to leptonic
 number violation, a clear signature for intermediate Majorana
 neutrino contribution. Such a signal (final dileptons of the same sign) is not
 possible if the heavy
 neutrinos are Dirac particles.  We present our results for the total
 cross-section as a function of the neutrino mass and the center of mass energies.}


\vspace{0.2in}
\vfill

\end{titlepage}
\vfill\eject

\narrowtext
\section{\bf Introduction}

Massive neutrinos can come in two different types: as Dirac or Majorana
particles. Dirac fermions have distinct particles and antiparticle degrees of
freedom while Majorana fermions make no such distinction and have half as many
degrees of freedom \cite{neutmass}. In this conditions fermions with conserved
charges (color, electric charged, lepton number,...) must be of Dirac type,
while fermions without conserved charges may be of either type.  New neutrinos
could have large masses and be of either type. If there are heavy neutrinos
(N), then the present and future experiments offer the possibility of
establishing their nature. The production of Majorana neutrinos via $e^+e^-$,
$e^- \gamma$, $\gamma \gamma$ and hadronic collision have been extensively
investigated in the past
\cite{ma,datta,gluza,hoefer,cevetic,almeida,jperessu,belanger}. In this work we
study the possibility of the $\gamma \gamma$ linear collider to produce clear
signatures for Majorana neutrinos. The photon linear collider \cite{ohgaki} may
be the best alternative to the electron positron colliders. In this paper we
discuss the possible detection of Majorana neutrinos in the reactions $\gamma
\gamma \rightarrow W^\pm W^\pm l_j^{\mp} l_k^{\mp}$ ($l_j\equiv e,\mu,\tau$)
where the real gamma beam are obtained by Comptom backscattering of laser
photons off the linear electron beam. In this process most of the photons are
produced at the high energy regions. For the cross section calculation we have
used the helicity formalism \cite{Causmaecker,stirling}, considering the
complete set of Feynman diagrams show in Fig.1. The phase space integration was
done taking account that the intermediate neutrinos can be either on shell or
off shell. In this point the present work differ of previous calculation
\cite{belanger} where the restriction $M_N>\sqrt{s}$ is considered and then
only neutrinos off-shell are take into account. Of course, for $M_N<\sqrt{s}$
the new heavy neutrino is more likely to be first discovered via single
production in $e^+e^- \rightarrow \nu N$. However the $\gamma\gamma$ collider
have some facilities that make it suitable to search Majorana neutrinos and
study their interactions. In particular a $\gamma\gamma$ collider offer the
possibility to control the helicity of the colliding photons controlling the
helicity of the Laser photons. Using these polarized high energy photon beams
we have the possibility to study in detail the interactions of Majorana
neutrinos and reject possible backgrounds. Due the large CM energies of these
colliders we have considered that the mass of the final lepton vanishes.

For the couplings of the Majorana neutrinos we follow Ref \cite{cevetic}
starting with rather general lagrangian densities for the interaction of $N$
with $W$ and light leptons $l_j$ ($e,\mu,\tau$):
\begin{equation}
{\cal L}_{NWl}=-\sum_{j=1}^3 \frac{g B_L^{(j)}}{\sqrt{2}} \bar l_j
\gamma^{\mu} P_L N W^-_{\mu} + h.c.
\end{equation}
The heavy Majorana neutrino couples to the three flavors lepton with couplings
proportional to $B_L^{(j)}$, where $j$ labels the family. This $(B_L^{(j)})^2$
constant are, in notation of \cite{nardi}, the sin square of the mixing angle
between light and heavy neutrinos ($(S_L^{\nu_e})^2$, $(S_L^{\nu_\mu})^2$,
$(S_L^{\nu_\tau})^2$). The constant $g$ is the standard $SU(2)_L$ gauge
coupling.

This $B_L^{(j)}$ parameters will affect the final results via the combinations
\begin{equation}
H_i=\mid B_L^{(i)}\mid^2 \mbox{\hspace{2cm} and \hspace{2cm} }
H=\sum_{j=1}^3\mid B_L^{(j)} \mid^2
\end{equation}
The final leptons can be either of $e^-$, $\mu^-$ or $\tau^-$ because this is
allowed by the interaction lagrangian (eq.1). All these possible final states
are a clear signal for intermediary Majorana neutrino and then we sum the cross
section over the flavors of the final lepton. This sum depend of $H$, $H_1$,
$H_2$ and $H_3$ in the numerator of the total cross section. In the other hand
this cross-section also depends on $H$ through the total width
$\Gamma_{N\rightarrow all}$ in the Majorana neutrino propagator. This last
dependence is relevant when the Majorana neutrino is on-shell. We use the
Breit-Wigner propagator for the Majorana neutrino $N$ for different values of
the mass $M_N$. The total width $\Gamma_{N \rightarrow all}$ of $N$ was
determined at tree level considering the dominant decay modes $N\rightarrow
W^{\pm}l_j^{\mp}$:

\begin{equation}
\Gamma_{N \rightarrow all}=\frac{g^2 H}{(32 \pi M_N^3 M_W^2)}
(M_N^2-M_W^2)(M_N^4+M_N^2 M_W^2-2 M_W^4)
\end{equation}

The constants $H_i$ ($i=1,2,3$) are severely restricted by available
experimental data (CERN $e^+e^-$ collider (LEP) and low energies data)
\cite{nardi,langacker}. This bounds are
\begin{eqnarray}
H_1 &<& 0.0066 \nonumber \\ H_2 &<& 0.006  \\ H_3 &<& 0.018
\nonumber
\end{eqnarray}
In this conditions the upper bound for $H$ is $H < 0.031$. However the bound on
$H_1$ is even more restricted once the constraint from $\beta\beta_{0\nu}$ are
taken into account \cite{balysh}. It is $H_1 < 2 \times 10^{-5}$ for $M_N < 1 $
TeV and then the electronic channel do not contribute appreciablement in the
kinematic region examined in the present work. The above mentioned bound do not
constraint the muonic and tauonic channel and do not put new limits on the
constants $H_2$ and $H_3$. Then, when the bounds from $\beta\beta_{0\nu}$ are
considered the new upper bound on $H$ is

\begin{equation}
H < 0.024.
\end{equation}

It is the value we have considered in the present article. In this work we have
considered the complete set of Feynman diagrams (Fig.1) that contribute at tree
level to $\gamma \gamma\rightarrow W^\pm W^\pm l^\mp_j l^\mp_k$ ($\rightarrow
jets + l_j l_k$) with the light leptons $l_1=e,l_2=\mu,l_3=\tau$.


 In $\mid \bar M \mid^2$ we average over the
initial photon polarizations and sum over the final polarization
of the both $W$ and $l^-_j$ and over the flavor of the final
leptons. Moreover an $\frac12$ factor is included to avoid double
counting of the two $W$ and another $\frac12$ to avoid double
counting of the two identical leptons when integrating over the
phase space. We take as inputs the values of $\sqrt{s}$ and $M_N$.
The $H$ dependence is complicate due to the Majorana neutrino
propagator . We have founded that the most important contribution
to $\mid \bar M \mid^2$ coming from the couplings is proportional
to $H^2$. However in the $M_W<M_N<\sqrt{s}-M_W$ kinematic region
(Reg.I), where the intermediate Majorana neutrino may be on-shell,
the total cross-section is approximately proportional to the $H$
value. In the other hand in the $M_N>\sqrt{s}-M_W$ region
(Reg.II), where the Majorana neutrino is off-shell, the total
cross-section is approximately proportional to $H^2$. The
behaviour in Reg.I is easy to realize if we make the so-called
peaking approximation, in which the Breit-Wigner shape of the
Majorana neutrino propagator is replaced by a delta function. In
this region the $H^2$ dependence in the numerator is canceled by
the $H$ factor in the total width. Considering only the relevant
factors in the cross section, we have

\begin{eqnarray}
\sigma=\sum_j \sigma_j \sim \cdots H^2
\frac{1}{(q^2-M_N^2)^2+M_N^2\Gamma_N^2}\cdots
\end{eqnarray}

where $j$ labels the final lepton flavors. Making now the peaking
approximation

\begin{eqnarray}
\cdots \frac{1}{(q^2-M_N^2)^2+M_N^2\Gamma_N^2}\cdots \rightarrow
\frac{\pi}{M_N \Gamma_N} \delta(q^2-M_N^2)
\end{eqnarray}

and since that $\Gamma \sim H$ (eq.3) then we can see that $\sigma=\sum_j
\sigma_j$ is approximately proportional to the $H$ value in Reg.I. The
expression for $M$ is too long. It extend over tens of page when printed out
and we do not present explicitly it here. We show the numerical results for the
cross-section for the $\gamma\gamma$-subprocess for different values of
$\sqrt{\widehat{s}}$ (the  center of mass energie of the $\gamma\gamma$
subprocess) and $M_N$. The numerical Monte Carlo integration was done using the
FORTRAN subroutine RAMBO \cite{rambo}.
 The Fig.2 show the $M_N$ dependence of the
unpolarized cross-section $\widehat{\sigma}$ at fixed $\sqrt{\widehat{s}}$ for
the maximum value of $H$ given by eq.5. To check the correctness of our final
4-body calculation we include in the same figure the corresponding results for
the final 3-body process ($\gamma\gamma\rightarrow \l^{\mp} N W^{\pm}$), for
$\sqrt{\widehat{s}}=500$ GeV. Moreover, partial results were also checked using
the CalcHep program \cite{calchep}. In Fig.3 we show the $\sqrt{\widehat{s}}$
dependence of $\widehat{\sigma}$ for different values of $M_N$ keeping the same
values for $H$.

In the following the embedding of the subprocess $\gamma \gamma \rightarrow
W^\pm W^\pm l_j^{\mp} l_k^{\mp}$ into appropriate photon fluxes is presented.
We have considered real gamma beam obtained by the Comptom backscattering of
laser photon off the linear electron beam. The mentioned subprocess is related
to $e^+ e^-$ collition by folding the $\gamma\gamma$-subprocess cross-section
($\widehat{\sigma}(\widehat{s})$) with appropriate $\gamma \gamma$ luminosity
function:

\begin{equation}
\sigma(s)=\int_{\tau^{min}}^{\tau^{max}} d\tau \frac{dL_{\gamma\gamma}}{d\tau}
\widehat{\sigma}_{\gamma \gamma \rightarrow W^{\pm} W^{\pm} l^{\mp}
l^{\mp}}(\tau s)
\end{equation}

The differential luminosity $dL_{\gamma\gamma}/d\tau$ is defined as usual in
terms of photons structure functions of electrons ($f(x)$) beam:

\begin{equation}
\frac{dL_{\gamma\gamma}}{d\tau}=\int_{\frac{\tau}{x^{max}}}^{x^{max}} \frac{d
x}{x} f(x) f(\frac{\tau}{x})
\end{equation}

where $\tau=\widehat{s}/s$. $\sqrt{\widehat{s}}$ is the center of
mass energy of the subsystem and $\tau^{min}$ is determined by the
production threshold:

\begin{equation}
\tau^{min}=(2 m_W)^2/s
\end{equation}

and $x$ is the fraction of the parent electron's energies carried
by the photons. The maximum possible value for $x$ is

\begin{equation}
x^{max}=\frac{\xi}{1+\xi} \mbox{,\hspace{2cm}where\hspace{2cm} } \xi=\frac{4
E_0 \omega_0}{m_e} ;
\end{equation}

where $E_0$ and $\omega_0$ are the incident electron and laser light energies.
To avoid unwanted $e^+e^-$ pair production from the collitions between the
incident and back-scattered photons, we should no choose too large $\omega_0$.
This fact constraint the maximum value for $\xi$ to be $\xi=2(1+\sqrt{2})$. The
variable $\tau$ is obviously $\tau=x_a x_b$, the product of electron's energies
fractions. In this conditions the minimal value for $x$ is $\tau/x^{max}$ and
the maximum value for $\tau$ is $(x^{max})^2$. This justify the integral limits
in eq(8) and (9).

In the other hand we can translate eq(8) and (9) in a alternative expression
for the cross-section:

\begin{equation}
 \sigma(s)=\int_{x_a^{min}}^{x^{max}}dx_a
\int_{\frac{\tau^{min}}{x_a}}^{x^{max}} dx_b f(x_a) f(x_b)
\widehat{\sigma}(x_a x_b s)
\end{equation}

where $x_a^{min}=\tau^{min}/x^{max}$.

To illustrate the possible impact of this process in the discovery of Majorana
neutrinos we show in Fig 4 the number of events as a function of $\sqrt{s}$ for
$M_N=300 GeV$.  We have considered the maximum possible value of $H$ as
inferred of the experimental bounds (eq.(5)).
 We have used the estimated luminosity
\cite{tesla} for the $\gamma \gamma$ collider of ${\cal L} = 100 fb^{-1}$.  We
have ignored the experimental difficulties of detecting the discussed process
unambiguously but if we take as reasonable the threshold of $100$ events then
we could see signatures for Majorana neutrinos with mass of $300$ GeV for
center of mass energies greater than $700$ GeV. We have also include in Fig.5 a
contour plot for different number of events as a function of $H$ and $\sqrt{s}$
for $M_N=300$ GeV. The horizontal line represent the maximum possible value for
$H$.

Summarizing, we calculate the cross-section for the process
$\gamma \gamma \rightarrow W^\pm W^\pm l_j^\mp l_k^\mp$ where
$l_1$, $l_2$ and $l_3$ are light leptons $(e,\mu,\tau)$
respectively. We show the total unpolarized cross-section for the
subprocess $\gamma\gamma$ for different values of $\sqrt{s}$ and
$M_N$. We have included all the contributions considering that the
intermediate Majorana neutrinos can be either on-shell or
off-shell. Finally we show the number of events as a function of
$\sqrt{s}$ considering the real gamma beam is obtained by the
Comptom backscattering of laser photons off the linear electron
beam.

{\bf Acknowledgements}

We thank CONICET (Argentina) and Universidad Nacional de Mar del
Plata (Argentina) for their financial supports.

\pagebreak

\noindent{\large \bf Figure Captions}\\

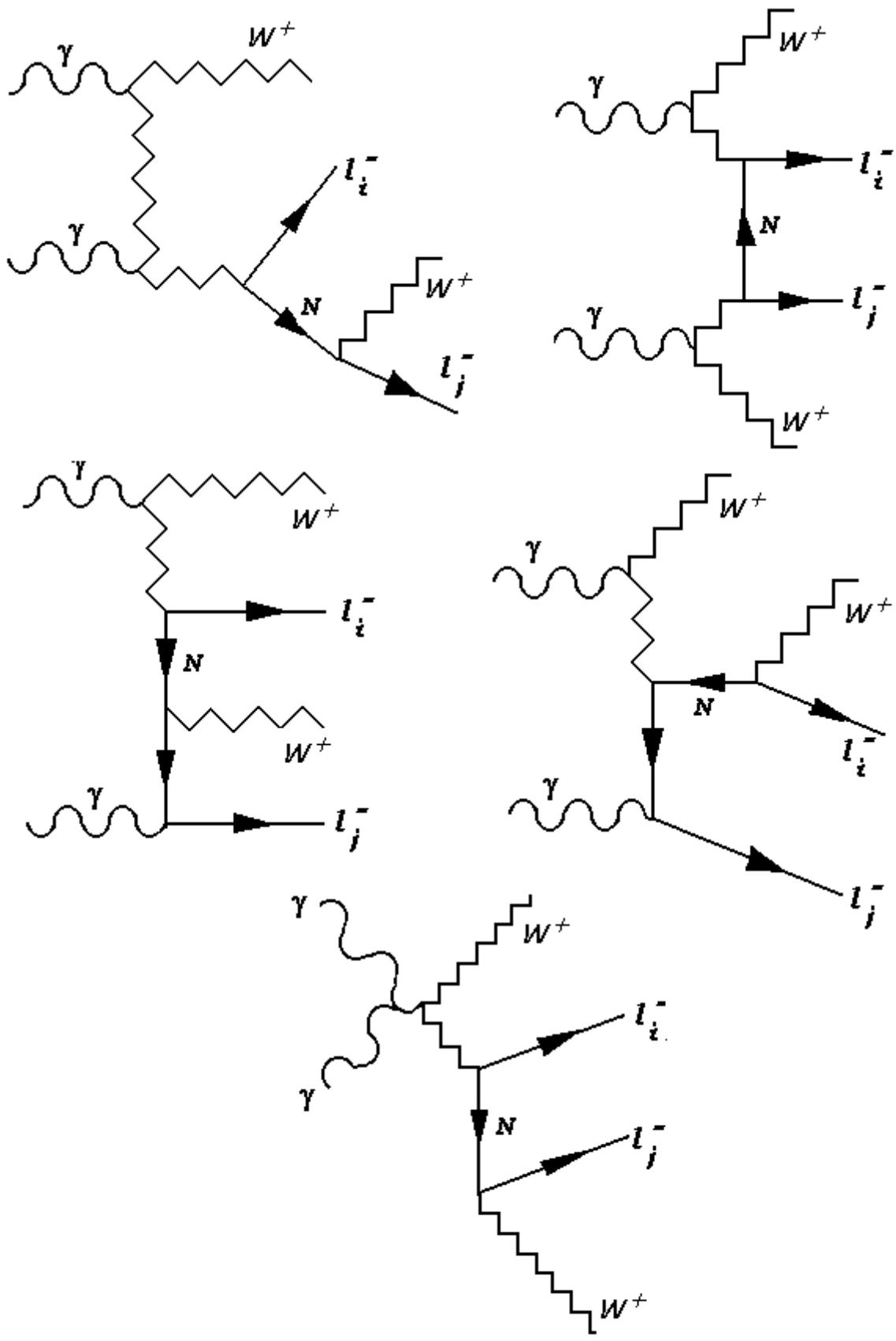
\noindent{\bf Figure 1:} Feynman graph contributing to the
amplitude of the $\gamma\gamma \rightarrow W^+W^+ l^- l^-$
process.

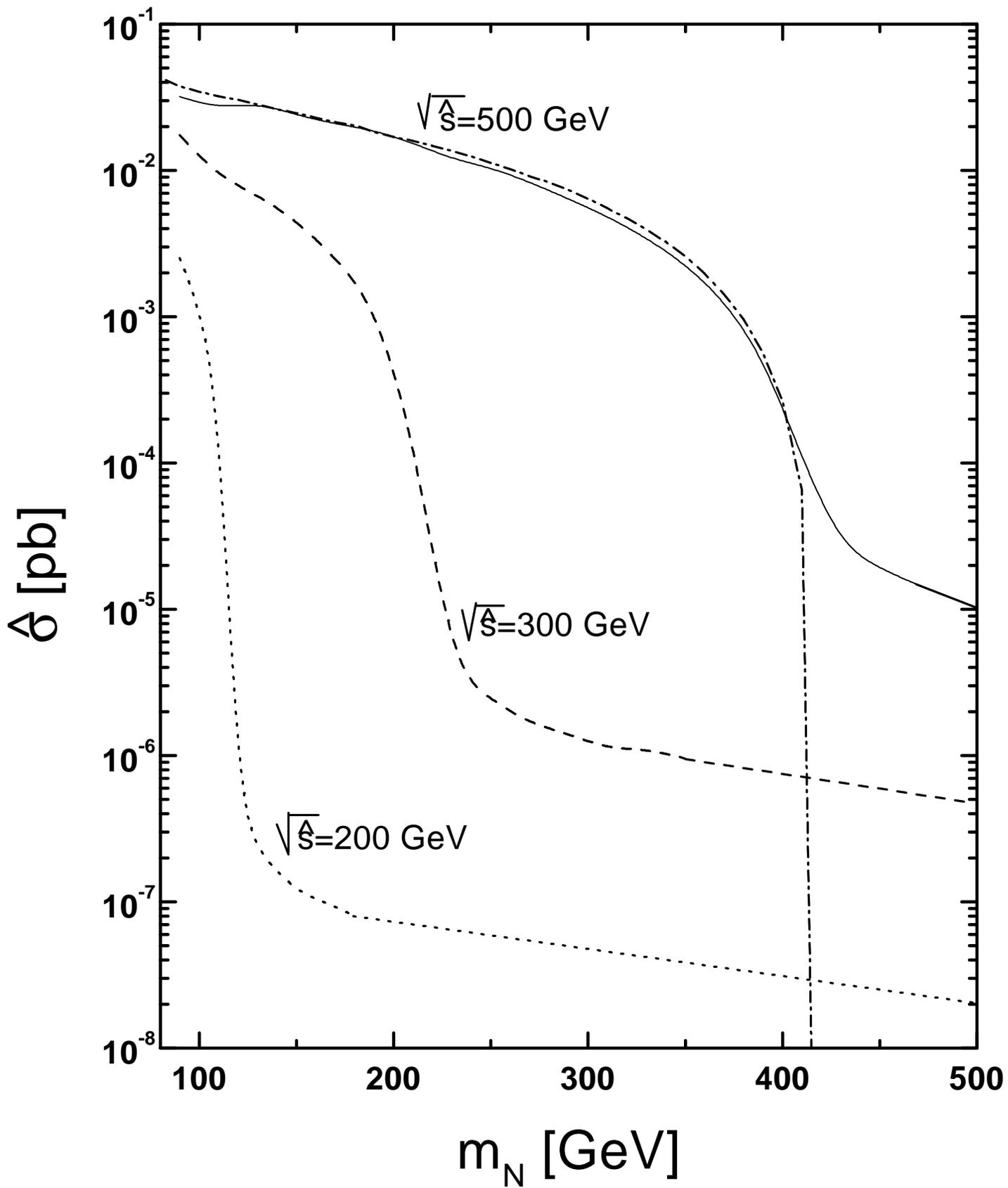
\noindent{\bf Figure 2:}Unpolarized cross-section ($\widehat{\sigma}$) for the
$\gamma\gamma$-subprocess as a function of the Majorana neutrino masses ($M_N$)
for different center of mass energies ($\sqrt{\widehat{s}}=$200, 300 and 500
GeV). To check the correctness of our final 4-body calculation we include, for
$\sqrt{\widehat{s}}=$500 GeV, a plot of $\sigma(\gamma\gamma\rightarrow l^{\mp}
N W^{\pm}) Br(N\rightarrow l^{\mp} W^{\pm})$ (Dot-Dash).

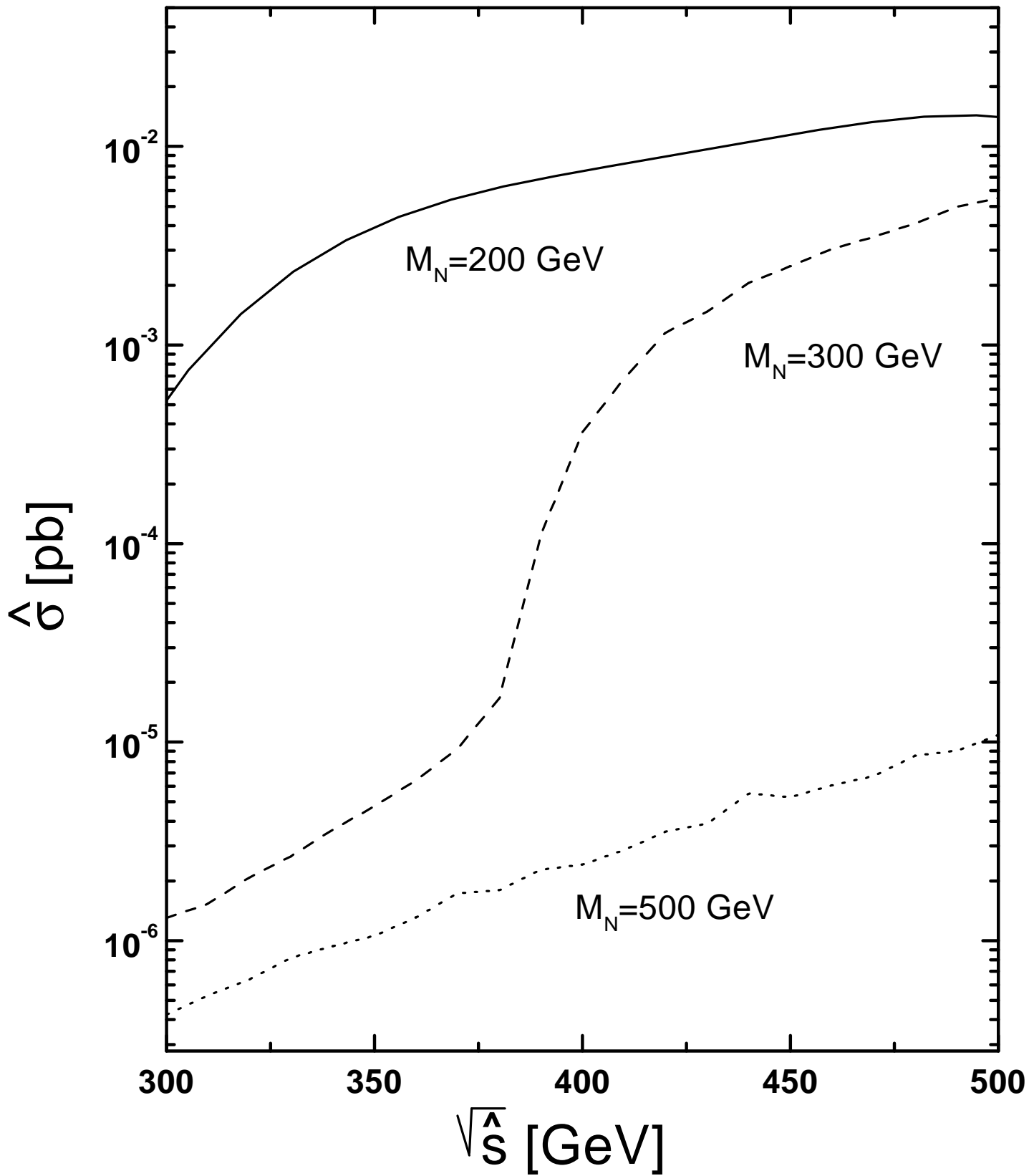
\noindent{\bf Figure 3:} Unpolarized cross section
($\widehat{\sigma}$) for the $\gamma\gamma$-subprocess as a
function of the center of mass energies ($\sqrt{\widehat{s}}$) for
different Majorana neutrino masses ($M_N$=200, 300 and 500 GeV).

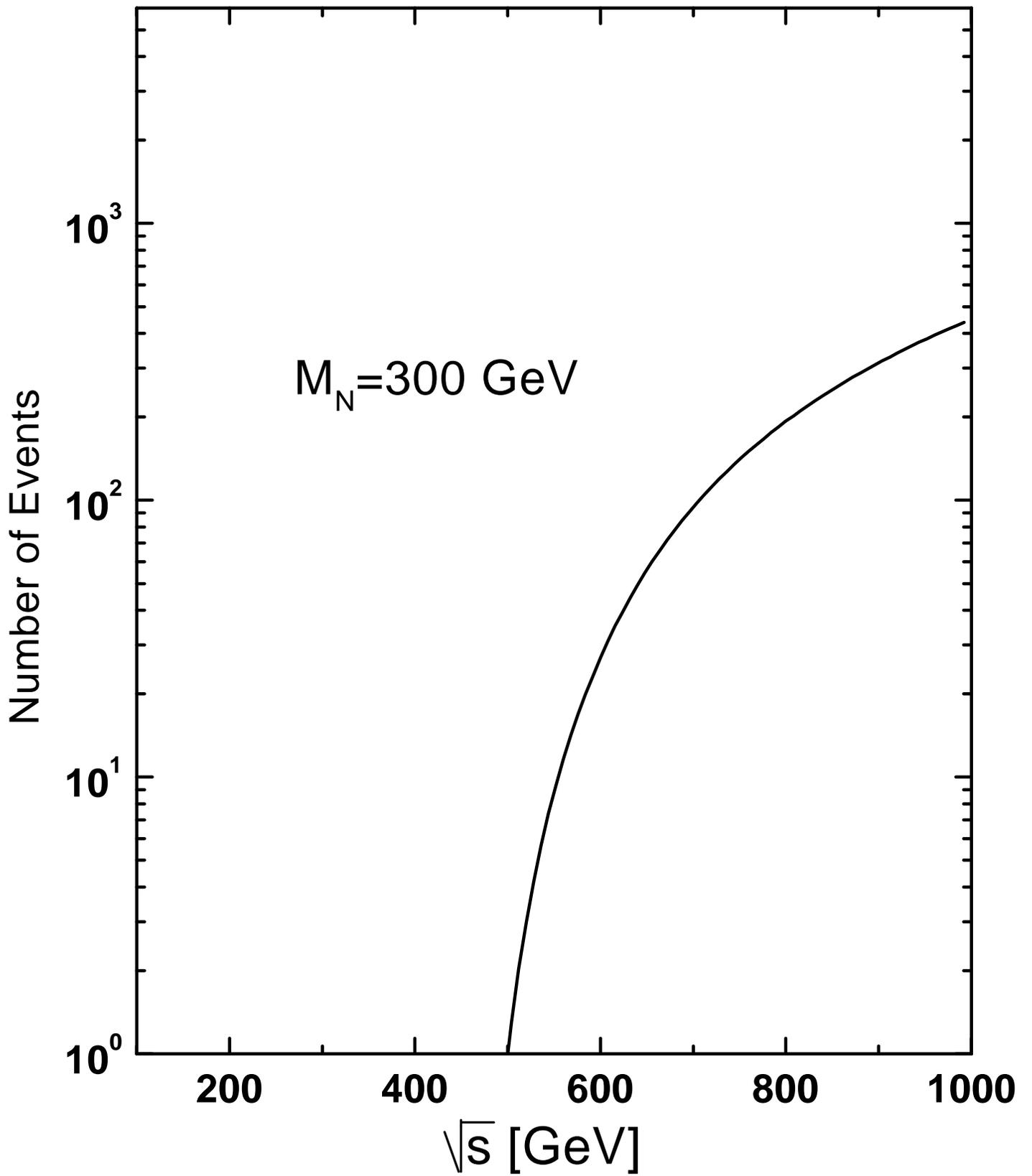
\noindent{\bf Figure 4:} Number of events as a function of
$\sqrt{s}$ for $M_N=300$ GeV.

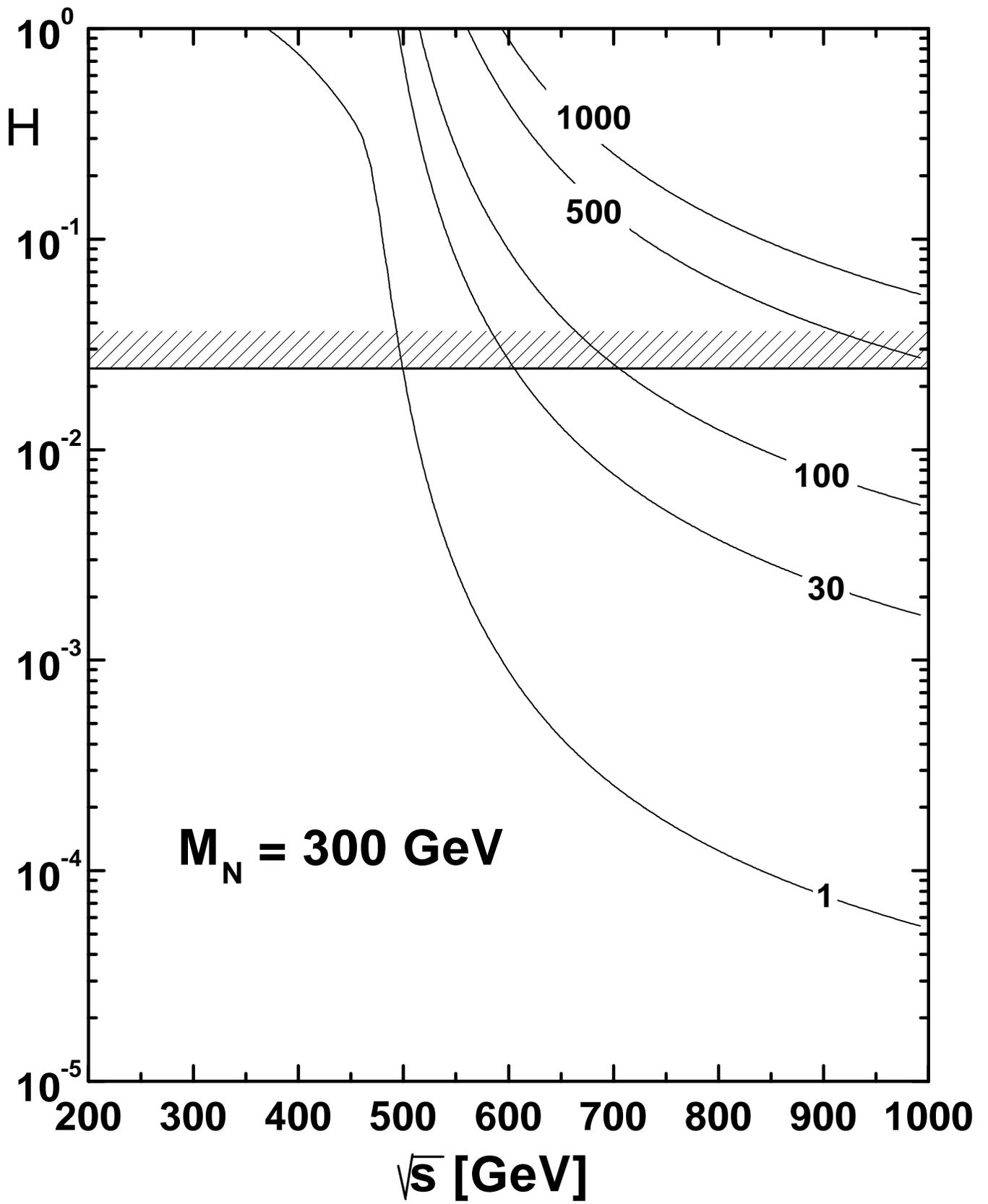
\noindent{\bf Figure 5:} Contour Plot for the number of events (1000, 500, 100,
30, 1) as a function of $\sqrt{s}$ and $H$. The horizontal line represent the
maximum possible value for $H$.

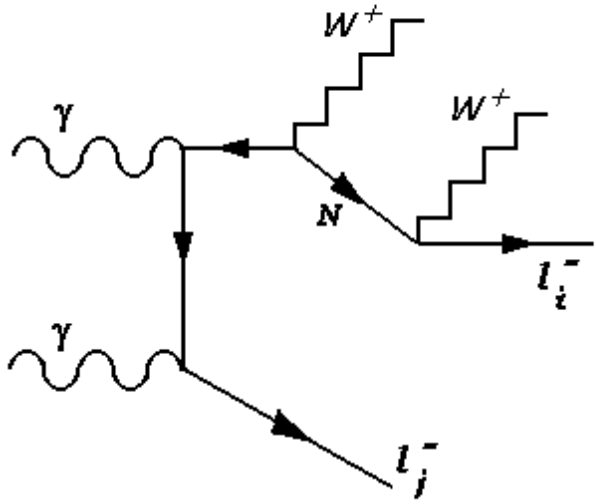
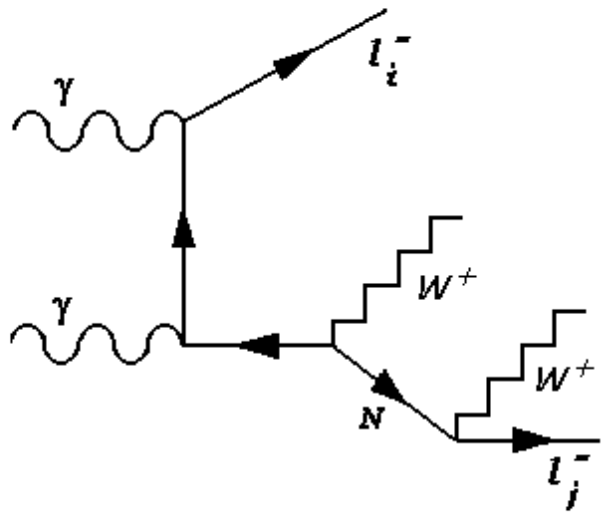
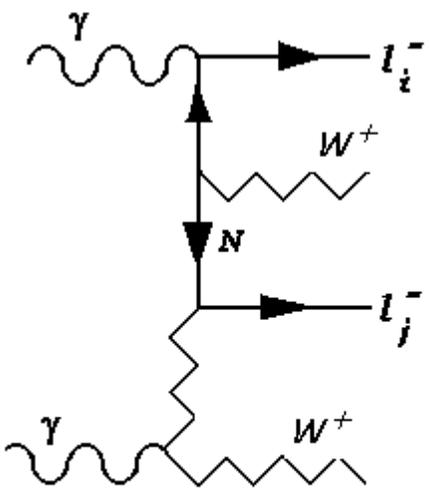
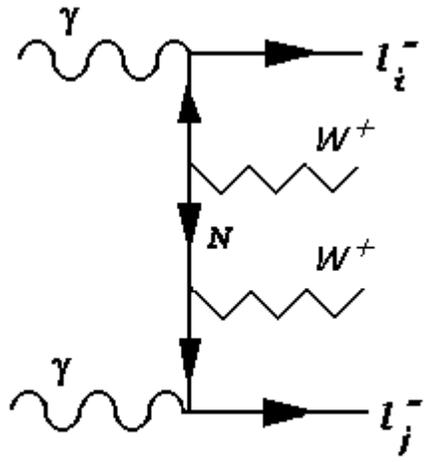
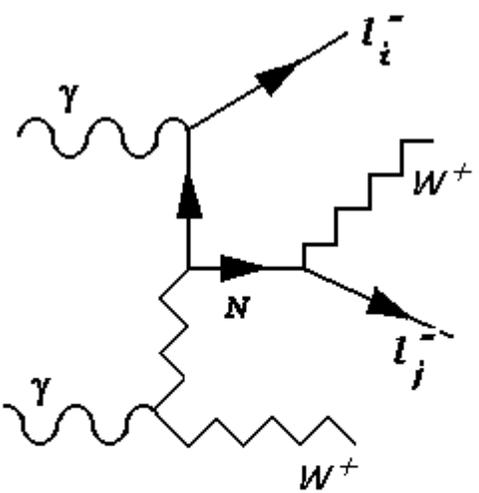
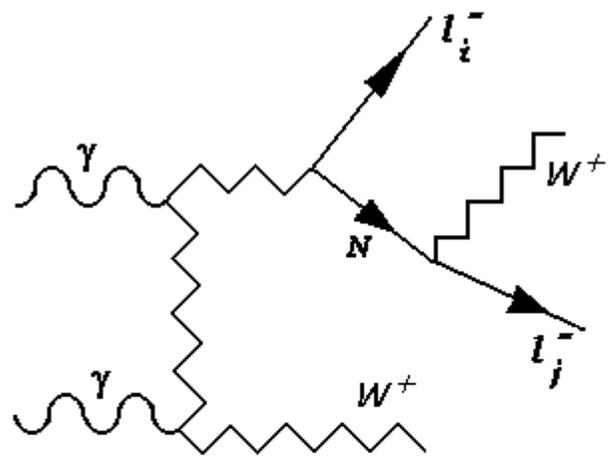
\end{document}